\documentclass[twocolumn,amsmath,amssymb,10pt,superscriptaddress,a4paper,letterpaper,fleqn]{revtex4-1}
\usepackage{amssymb}
\usepackage{epsfig}
\usepackage{graphicx}
\usepackage{dcolumn}
\usepackage{array}
\usepackage{bm}
\usepackage{fancyheadings}
\usepackage{longtable}
\usepackage{multirow}
\usepackage{float}
\usepackage{afterpage}  
\usepackage{color}
\usepackage{array}

\setlongtables
\usepackage[breaklinks=true,linkbordercolor={1 1 1}]{hyperref}

\begin{document}
\setcounter{page}{1}


\title{
\qquad \\ \qquad \\ \qquad \\  \qquad \\  \qquad \\ \qquad \\ 
B(E2) Evaluation for 0$^{+}_{1}$ $\rightarrow$ 2$^{+}_{1}$ Transitions in Even-Even Nuclei
}
\author{
B. Pritychenko
}
\email[Corresponding author, electronic address:\\ ]{pritychenko@bnl.gov}
\affiliation{
National Nuclear Data Center, Brookhaven National Laboratory, Upton, NY 11973-5000, USA
}
\author{
M. Birch
}
\affiliation{
Department of Physics \& Astronomy, McMaster University, Hamilton, Ontario L8S 4M1, Canada
}
\author{
M. Horoi
}
\affiliation{
Department of Physics, Central Michigan University, Mount Pleasant, MI 48859, USA
}
\author{
B. Singh
}
\affiliation{
Department of Physics \& Astronomy, McMaster University, Hamilton, Ontario L8S 4M1, Canada
}
\date{\today} 

\begin{abstract}
{
A collaborative study by Brookhaven-McMaster-Central Michigan  is underway to evaluate B(E2)$\uparrow$ for 0$^{+}_{1}$ $\rightarrow$ 2$^{+}_{1}$ transitions. This work is a continuation of a previous USNDP evaluation and has been motivated by a large number of recent measurements and nuclear theory developments. It includes an extended compilation, data evaluation procedures and shell model calculations. The subset  of B(E2)$\uparrow$ recommended values for nuclei of relevance to the double-beta decay problem is presented, and   evaluation policies of experimental data and systematics are discussed. Future plans for completion of the B(E2;0$^{+}_{1}$ $\rightarrow$ 2$^{+}_{1}$) evaluation project are also described. 
}
\end{abstract}
\maketitle


\lhead{ND 2013 Article $\dots$}
\chead{NUCLEAR DATA SHEETS}
\rhead{B. Pritychenko, M. Birch, ...}
\lfoot{}
\rfoot{}
\renewcommand{\footrulewidth}{0.4pt}


\section{ INTRODUCTION}

The importance of B(E2)$\uparrow$ values compilation and evaluation was recognized in the 1960s by Stelson  and Grodzins  \cite{65St,62Gro}. 
Evaluated quadrupole collectivity (reduced electric quadrupole transition rates)  values are essential for  nuclear structure physics and in 
high demand for nuclear model calculations. These nuclear data activities flourished at  Oak Ridge National Laboratory at the end of the century under the leadership of Raman, who produced two excellent evaluations \cite{87Ram,01Ram}. In recent years, this project was transferred to Brookhaven National Laboratory in order to ensure continuation of this important work. 

The volume of B(E2)$\uparrow$ data is rapidly increasing due to the general availability of supercomputers and nuclear radioactive beam facilities. These
facilities have been producing rare nuclei far from the valley of stability at an increasing rate,  providing researchers
with unprecedented opportunities to study their properties. In many cases, B(E2) values and energies of the low-lying
states have been studied for the first time. Large amounts of new data, especially for the A$\leq$100 region, require a new
evaluation of quadrupole collectivities for proper interpretation and analysis of the newly-obtained values.

Access to these   data is of paramount importance for  the SciDAC ad FRIB collaborations \cite{Sci,FRIB}. To satisfy the evolving needs of theoretical and experimental nuclear physics,    a new B(E2)$\uparrow$ evaluation has been launched \cite{12Pri}.  This work broadens the previous  evaluation  of even-even nuclei  \cite{01Ram} along the nuclear landscape, and includes many new experimental and evaluated quantities. It consists of an updated set of  compiled experimental parameters, comprehensive data analysis and evaluation and  large-scale shell model calculations. It extends the list of evaluated nuclei  from 53 to 68 and  from 20 to 38 in the Z=2-22  and   Z=24-30 regions, respectively. More details are given in the following sections.

\begin{figure}[!htb]
 \begin{center}
\fbox{\includegraphics[height=4cm]{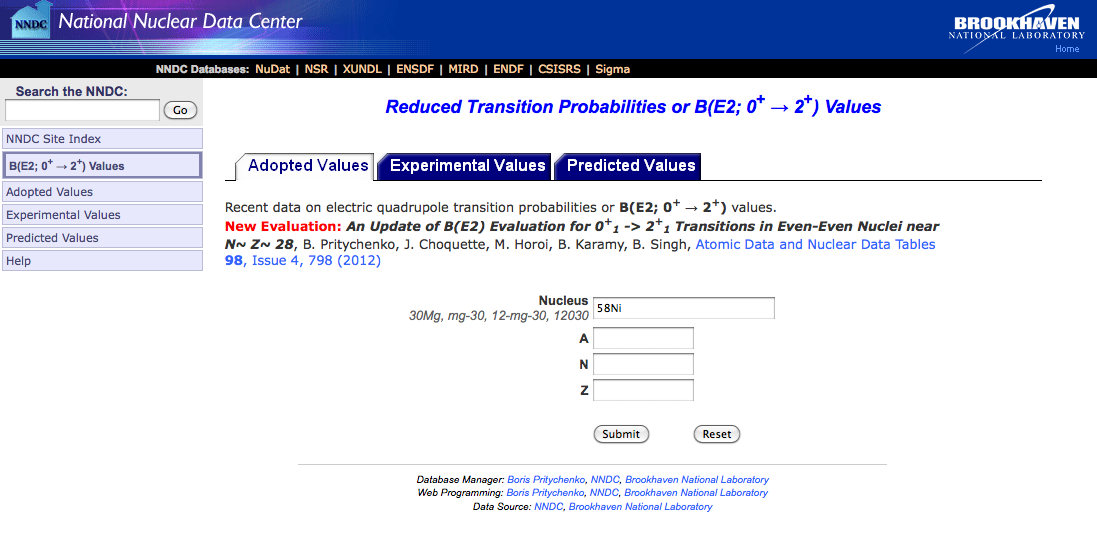}}
\caption{B(E2)$\uparrow$  Web Interface  http://www.nndc.bnl.gov/be2. Example of evaluated data search for $^{58}$Ni.}
\label{fig1}
\end{center}
\end{figure}

\renewcommand{\arraystretch}{0.2}
\begin{table*}[!htb]
\centering
\caption{Recommended B(E2;0$^{+}_{1}$ $\rightarrow$ 2$^{+}_{1}$) and $\beta_2$  values for 2$\beta^-$ candidates. }
\begin{tabular}{c|c|c|c|c|c|c|c}
\hline
\hline
Parent  &  E(2$^{+}_{1}$), keV & B(E2)$\uparrow$, W.u. &  $\beta_2$ &   Daughter  &  E(2$^{+}_{1}$), keV & B(E2)$\uparrow$, W.u.  &  $\beta_2$ \\
\hline
$^{46}$Ca & 1346.0 (3) & 3.61(+32-15) & 0.1505(+66-31) & $^{46}$Ti  & 889.286 (3) & 19.42(52) & 0.3175(42) \\ 
$^{48}$Ca & 3831.72 (6) & 1.768(+234-96) & 0.1054(+70-29) & $^{48}$Ti  & 983.5390 (24) & 12.77(56) & 0.2575(56) \\ 		
$^{70}$Zn & 884.46(8) & 17.80(88) & 0.2229(55) & $^{70}$Ge  & 1039.485(22) & 19.61(77) & 0.2194(43) \\ 
$^{76}$Ge & 562.93(3) & 28.16(+79-76) & 0.2629(+37-36) & $^{76}$Se  & 559.102(5) & 45.14(+157-59) & 0.3133(+55-20) \\ 	
$^{80}$Se & 666.27(7) & 24.62(80) & 0.2314(38) & $^{80}$Kr  & 616.60(10) & 37.2(11)& 0.2686(41) \\ 
$^{82}$Se & 654.75(16) & 17.3(9) & 0.1939(53) & $^{82}$Kr  & 776.520(3) & 21.28(+63-58) & 0.2031(+30-28) \\ 	
$^{86}$Kr & 1564.75(10) & 9.36(84) & 0.1347(60) & $^{86}$Sr  & 1076.68(4) & 11.89(68) & 0.1439(41) \\ 
$^{94}$Zr & 918.75(5) & 4.96(35) & 0.0882(31) & $^{94}$Mo  & 871.098(16) & 16.32(58) & 0.1525(27) \\ 	
$^{96}$Zr & 1750.497(15) & 2.38(32) & 0.0611(41) & $^{96}$Mo  & 871.098(16) & 16.32(58) & 0.1525(27) \\ 
$^{98}$Mo & 787.384(13) & 20.09(43) & 0.1692(18) & $^{98}$Ru  & 652.44(4) & 29.89(97) & 0.1970(32) \\ 		
$^{100}$Mo & 535.561(22) & 38.4(16) & 0.2340(49) & $^{100}$Ru  & 539.510(10) & 35.74(30) & 0.21539(90) \\ 
$^{104}$Ru & 358.02(7) & 56.9(12) & 0.2717(28) & $^{104}$Pd  & 555.81(4) & 36.4(10) & 0.2080(29) \\ 	
$^{110}$Pd & 373.80(6) & 55.3(15) & 0.2562(34) & $^{110}$Cd  & 657.7645(20) & 27.20(68) & 0.1722(22) \\ 
$^{114}$Cd & 558.456(2) & 32.63(46) & 0.1886(13) & $^{114}$Sn  & 1299.907(7) & 12.69(67) & 0.1130(30) \\ 	
$^{116}$Cd & 513.490(15) & 34.51(89) & 0.1940(25) & $^{116}$Sn  & 1293.560(8) & 11.67(38) & 0.1083(18) \\ 
$^{122}$Sn & 1140.51(3) & 10.38(49) & 0.1021(24) & $^{122}$Te  & 564.094(16) & 36.18(+73-69) & 0.1834(18) \\ 
$^{124}$Sn & 1131.739(17) & 9.11(23) & 0.0957(12) & $^{124}$Te  & 602.7271(21) & 31.6(16) & 0.1713(45) \\ 
$^{128}$Te & 743.219(7) & 19.83(37) & 0.1358(13) & $^{128}$Xe  & 442.911(9) & 40.2(16) & 0.1862(37) \\ 		
$^{130}$Te &  839.494(17) &  15.21(36) &  0.1189(14) & $^{130}$Xe  & 536.068(6) & 30.8(+14-11) & 0.1630(+38-28) \\ 
$^{134}$Xe &  847.041(23) & 16.5(13) & 0.1192(48) & $^{134}$Ba  & 604.7223(19) & 32.63(95) & 0.1617(23) \\ 	
$^{136}$Xe & 1313.027(10) & 13.8(39) & 0.109(15) & $^{136}$Ba  & 818.497(11) & 19.87(+54-53) & 0.1262(17) \\ 
\hline
\hline
\end{tabular}
\label{table1}
\end{table*}

\section{COMPILATION AND EVALUATION OF EXPERIMENTAL DATA}
\label{sec:evaluation}
 
The quadrupole collectivity  data compilation at Brookhaven was accompanied by the  online  
service  http://www.nndc.bnl.gov/be2  \cite{06Pri} and later evolved into a collaborative data evaluation project. 
The B(E2)$\uparrow$ project  Web Interface is shown in Fig. \ref{fig1}. 

\begin{figure}[!htb]
 \begin{center}
\fbox{\includegraphics[height=6cm]{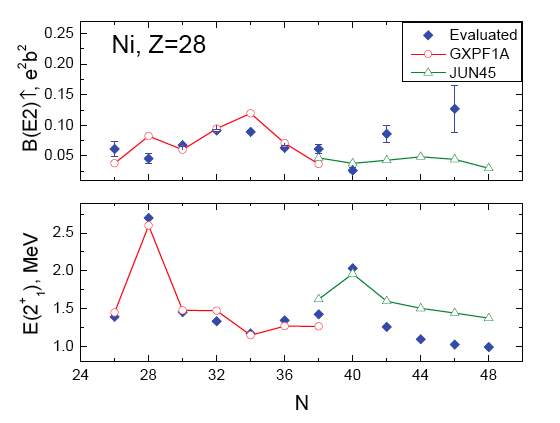}}
\caption{Brookhaven-McMaster-Central Michigan evaluated B(E2)$\uparrow$ values and 2$^{+}_{1}$  energies vs. shell-model calculations for Ni  \cite{12Pri}.}
\label{fig2}
\end{center}
\end{figure}

This compilation is based on the previous evaluation \cite{01Ram},    XUNDL and NSR database content \cite{xundl,13Pri} 
and the original research papers.  All measurements were separated into three classes:   
model independent, low model dependent and model dependent \cite{01Ram,12Pri}. 
The recommended B(E2)$\uparrow$ values were deduced using model-independent
or traditional, combined (model-independent and low model-dependent),
and model-dependent datasets with the AveTools
software package \cite{10Ave} using the selected datasets. The final results were analyzed using the shell model predictions. 
A subset of evaluated and calculated data for Ni nuclei is shown in Fig. \ref{fig2}.

 As an example, consider a non-tradional application of B(E2) $\uparrow$ evaluation. In recent years a considerable amount of effort has been dedicated to the  search for double-beta decay transitions to the first excited states \cite{02Tre}. 
 Two-neutrino double-beta decay half-lives for the transitions to the first excited 2$^+$ and 0$^+$ states are often computed on the basis of a second Quasi Random Phase Approximation (QRPA) and Hartree-Fock-Bogoliubov models \cite{95Sto,04Rai}.  The deformed QRPA using the realistic Bonn-CD nucleon-nucleon interaction approach has been applied recently for the neutrinoless mode \cite{12Fan}. The reliability of these calculations has been tested by comparing  theoretically calculated results for a number of spectroscopic properties such as reduced  transition probabilities  with evaluated data.   
To facilitate such calculations, two subsets of evaluated  nuclear structure data for 2$\beta^-$-decay and 2$\beta^+$-, $\epsilon$$\beta^+$-  and 2$\epsilon$-decay candidates \cite{92Boe} are shown in Tables \ref{table1} and \ref{table2}, respectively.  
These tables  reflect   current progress and indicate future work.

\renewcommand{\arraystretch}{0.2}
\begin{table*}[!htb]
\centering
\caption{Recommended B(E2;0$^{+}_{1}$ $\rightarrow$ 2$^{+}_{1}$) and $\beta_2$  values for 2$\beta^+$, $\epsilon$$\beta^+$  and 2$\epsilon$ candidates. }
\begin{tabular}{c|c|c|c|c|c|c|c}
\hline
\hline
Parent  &  E(2$^{+}_{1}$), keV & B(E2)$\uparrow$, W.u. &  $\beta_2$ &   Daughter  &  E(2$^{+}_{1}$), keV & B(E2)$\uparrow$, W.u.  &  $\beta_2$ \\
\hline
$^{50}$Cr & 783.3(9) & 19.32(42) & 0.2903(32) & $^{50}$Ti  & 1553.778 (7) & 5.04(30) & 0.1617(48) \\ 
$^{58}$Ni & 1454.21(9) & 10.04(17) & 0.1794(15) & $^{58}$Fe & 810.7662(20) & 16.9(24) & 0.250(18) \\ 		
$^{64}$Zn & 991.56(5) & 19.52(68) & 0.2335(41) & $^{64}$Ni  & 1345.75(5) & 8.3(5) & 0.163(5) \\ 
$^{74}$Se & 634.74(6) & 38.7(21) & 0.2902(80) & $^{74}$Ge & 595.850(6) & 32.68(+90-81) & 0.2832(+39-35)  \\ 
$^{78}$Kr & 455.033(23) & 64.0(16) & 0.3524(44) & $^{78}$Se  & 613.727(3) & 34.6(12) & 0.2744(49) \\ 
$^{84}$Sr & 793.22(6) & 26.7(21) & 0.2156(84) & $^{84}$Kr & 881.615(3) & 11.60(+44-25) & 0.1500(+29-16) \\ 		
$^{92}$Mo & 1509.51(3) & 7.90(35) & 0.1061(23) & $^{92}$Zr  & 934.47(5) & 6.49(32) & 0.1009(25) \\ 
$^{96}$Ru & 832.56(5) & 18.22(50) & 0.1538(21) & $^{96}$Mo & 778.237(10) & 21.26(45) & 0.1740(19) \\ 
$^{102}$Pd & 556.44(5) & 32.5(16) & 0.1965(49) & $^{102}$Ru  & 475.0962(10) & 44.68(88) & 0.2408(24) \\ 
$^{106}$Cd & 632.64(4) & 27.32(83) & 0.1726(26) & $^{106}$Pd & 511.850(23) & 44.3(12) & 0.2294(30) \\ 		
$^{112}$Sn & 1256.85(7) & 14.47(58) &  0.1206(24) & $^{112}$Cd  & 617.520(10) & 30.02(91) & 0.1810(27) \\ 
$^{120}$Te & 560.438(20) & 40.3(20) & 0.1936(47)  & $^{120}$Sn & 1171.265(15) & 11.23(29) & 0.1062(14) \\ 
$^{124}$Xe & 354.04(4) & 59.8(+24-21) & 0.2269(+46-40) & $^{124}$Te  & 602.7271(21) & 31.6(16) & 0.1713(45) \\ 
$^{130}$Ba &  357.38(8) & 58.2(24) & 0.2159(44) & $^{130}$Xe & 536.068(6) & 30.8(+14-11) & 0.1630(+38-28) \\ 		
\hline
\hline
\end{tabular}
\label{table2}
\end{table*}

\section{ CONCLUSIONS}
The B(E2)$\uparrow$ evaluation is proceeding under the auspices of the U.S. Nuclear Data Program. This effort is continuation of the previous works of  Stelson, Grodzins  and Raman \cite{65St,62Gro,87Ram,01Ram}.   This procedure  includes the  broadened nuclear structure data sets supported by shell model calculations for all known  even-even nuclei. 

The subset of the latest quadrupole collectivity data relevant to double-beta decay problem has been noted \cite{92Boe}. These well-established data could provide some guidance for theoretical calculations of double-beta decay rates for nuclei of interest where information on Gamow-Teller transitions is lacking. 

The data evaluation for the  Z$\sim$28 nuclei \cite{12Pri} is publicly available and the whole Z=2-56 region is completed. The evaluation of  the Z$>$56 region is currently underway and  will become available in the next few years.

{\it Acknowledgments}: 
We are indebted to M. Herman (BNL) for his support of this project,  and grateful to M. Blennau (BNL) for help with the manuscript. This work was funded by the Office of
Nuclear Physics, Office of Science of the U.S. Department of Energy, under Contract No. DE-AC02-98CH10886 with
Brookhaven Science Associates, LC. 

Work at McMaster University was  supported by DOE and NSERC of Canada. MH acknowledges support from DOE grant DE-FC02-09ER41584 (UNEDF SciDAC Collaboration).

\end{document}